# Which early works are cited most frequently in climate change research literature?

## A bibliometric approach based on Reference Publication Year Spectroscopy

Werner Marx*+, Robin Haunschild*, Andreas Thor**, Lutz Bornmann***

* Max Planck Institute for Solid State Research
Heisenbergstraße 1, 70569 Stuttgart, Germany
E-mail: w.marx@fkf.mpg.de, r.haunschild@fkf.mpg.de

** Hochschule für Telekommunikation Leipzig
Gustav-Freytag-Str. 43-45, 04277 Leipzig, Germany
E-Mail: thor@hft-leipzig.de

*** Division for Science and Innovation Studies
Administrative Headquarters of the Max Planck Society
Hofgartenstr. 8,
80539 Munich, Germany.
E-mail: bornmann@gv.mpg.de

+ Corresponding author

**Abstract**

This bibliometric analysis focuses on the general history of climate change research and, more specifically, on the discovery of the greenhouse effect. First, the Reference Publication Year Spectroscopy (RPYS) is applied to a large publication set on climate change of 222,060 papers published between 1980 and 2014. The references cited therein were extracted and analyzed with regard to publications, which are cited most frequently. Second, a new method for establishing a more subject-specific publication set for applying RPYS (based on the co-citations of a marker reference) is proposed (RPYS-CO). The RPYS of the climate change literature focuses on the history of climate change research in total. We identified 35 highly-cited publications across all disciplines, which include fundamental early scientific works of the 19[th] century (with a weak connection to climate change) and some cornerstones of science with a stronger connection to climate change. By using the Arrhenius (1896) paper as a RPYS-CO marker paper, we selected only publications specifically discussing the discovery of the greenhouse effect and the role of carbon dioxide. Also, we focused on the time period 1800-1850 to reveal the contributions of J.B.J Fourier in terms of cited references. Using different RPYS approaches in this study, we were able to identify the complete range of works of the celebrated icons as well as many less known works relevant for the history of climate change research. The analyses confirmed the potential of the RPYS method for historical studies: Seminal papers are detected on the basis of the references cited by the overall community without any further assumptions.

## 1. Introduction

Climate change has gained strongly increasing attention in the natural sciences and more recently also in the social and political sciences. Scientists actively work to understand past climate and to predict future climate by using observations and theoretical models. Various subfields from physics, chemistry, meteorology, and geosciences (atmospheric chemistry and physics, geochemistry and geophysics, oceanography, paleoclimatology etc.) are interlinked. The scientific community has contributed extensively with various data, discussions, and projections on the future climate as well as on the effects and risks of the expected climatic change (IPCC Synthesis Report, 2014).

Bibliometrics as a rapidly developing quantitative method is not only useful for research assessment purposes, but also e.g. for analyzing the history of science. This bibliometric analysis focuses on the history of climate change research: Which are the early works that are still alive in recent climate change related publications in the form of most frequently cited references and thus are most important for the evolution of this research field? On whose shoulders do the publishing authors stand and which are the origins or intellectual roots? Such questions can be answered by using a bibliometric method called Reference Publication Year Spectroscopy (RPYS, see Marx et al., 2014).

By putting climate change research more in the viewpoint of science history, we checked whether the recently highlighted and celebrated icons like Svante Arrhenius and Charles Keeling would be reflected in the results of a bibliometric approach and which additional works are cited most frequently. Science historians claim different works as signpost in the timeline of climate change and often cite them inconsistently, if at all (Fleming, 1998, 1999). Our bibliometric approach, however, is based on the literature selection by and accentuation of the overall climate change research community and thus shows a more complete and differentiated picture. The method allows us to identify the complete range of the works cited in the context of the history of climate change research and the discovery of the greenhouse effect, respectively.

RPYS is based on the assumption that peers produce a useful databasis by their publications, in particular by the references cited therein. This databasis can be analysed statistically with regard to the works most relevant for their specific field of research. Whereas individual scientists judge the origins of their research field more or less subjectively, the overall community might deliver a more objective picture. The peers effectively "vote" by their cited references, which works turned out to be most important for the evolution of their research field.

RPYS utilizes the following observation: The analysis of the publication years of the references cited by the papers in a specific research field shows that earlier publication years are not equally represented, but that some years occur particularly frequently among the references. These years appear in the distribution of the reference publication years as pronounced peaks. The peaks are usually based on single early publications, which were most frequently cited. These frequently cited papers are – as a rule – of specific significance to the research field in question and often represent its origins and intellectual roots (Marx et al., 2014). In recent years, several publications have appeared, in which the RPYS was described and applied to examine the origins of research fields (Marx and Bornmann, 2014; Barth et al., 2014; Leydesdorff et al., 2014; Comins and Hussey, 2015a, 2015b). Compared to these studies, our analysis is based on a very large publication and reference set.

Since only a small portion of the publications in a specific field of research normally discusses its historical background, a kind of dilution or weakening happens with regard to the appearance of historical references: Among the multitude of cited references, only the very prominent and most cited early works appear as distinct peaks in the RPYS. Therefore, we present in this study an extension of RPYS for the compilation of the publication set to be analyzed including the references cited therein (RPYS-CO). We apply a bibliometric method, which has been used extensively for many other purposes: the method of co-citation analysis (Small,1977).

This method takes advantage of the fact that concurrently cited (co-cited) papers are more or less closely related to each other. One can select the citation environment of a specific reference (or of two or even more references) in the form of all co-cited references and analyze these references applying RPYS-CO. The specific reference should be a prominent and seminal work which is used as a kind of marker or tracer reference for a specific topic in a field. We assume that papers which cite the selected reference(s) are potential candidates for citing also many other references relevant in a specific historical context. Thus, we select from the complete publication set a much smaller part focusing on a specific topic (here: of the discovery of the greenhouse effect and the specific role of carbon dioxide).

In this study we present the results of the RPYS and RPYS-CO on climate change research (1) based on the total number of references (n=10,932,050) cited within a carefully searched publication set of papers most relevant for this research field (n= 222,060) as well as (2) based only on the co-cited references of a single paper (n=1658). For RPYS-CO, we selected the paper by Svante Arrhenius (1896) entitled "On the influence of carbonic acid in the air upon the temperature of the ground" which has been cited around 600 times until present. Both analyses have been performed with the CRExplorer (www.crexplorer.net; Thor et al., 2016) – a program which has been designed for RPYS. This paper introduces the new option of the CRExplorer to undertake a RPYS-CO on a restricted set of cited references which are co-cited with at least one selected cited reference: the marker reference.

## 2. Datasets and Methodology
### 2.1 Climate change research literature in total

The analyses are based on the WoS custom data of our in-house database derived from the Science Citation Index Expanded (SCI-E), Social Sciences Citation Index (SSCI), and Arts and Humanities Citation Index (AHCI) produced by Thomson Reuters (Philadelphia, USA). First, we applied a sophisticated search method called "interactive query formulation" to compile the publication set of papers dealing with climate change (see Haunschild et al., 2016). The search strategy comprises a preliminary search for key papers and a renewed search based on the synonyms revealed by the keyword analysis of the key papers (Wacholder, 2011). The search was restricted to the time period 1980-2014 and resulted in a publication set of 222,060 papers (articles and reviews only).

This is not the complete publication set covering any research paper relevant for the climate change research field. However, we assume that we have included by far most of the relevant papers, in particular the key papers dealing with research on climate change. Furthermore, we expect that a moderate amount of incompleteness does not bias the RPYS. An expansion of the search query would increase the possibility of including non-relevant publications in the set.

For the RPYS, all cited references (n=10,932,050) have been selected from the papers in our publication set on climate change (n=222,060), i.e. around 50 references per paper. To cope with the large number of almost 11 million cited references when using the CRExplorer, we applied the following processing strategy:

1. We imported only references with publication years earlier than 1971 into the CRExplorer. This reduced the overall number of references substantially from originally 10,932,050 to 239,887 references. The limitation of the references to those with publication years prior to 1971 follows from the objective of this analysis to detect influential early works.

2. Reference variants of the same paper were algorithmically clustered and merged with the CRExplorer. Additionally, we applied manual clustering to the references published prior to 1900. For example, two variants of the paper by Arrhenius (1896) were merged, which has been cited 279 times with "Philosophical Magazine" and 32 times with "London Edinburgh Dublin" as journal title.

3. In order to detect the very early referenced publications, we selected the reference publication year time period 1000-1900 and removed all references with reference counts less than 10 for further analysis. The number of references decreased to n=152. A minimum reference count of 10 has proved to be reasonable for referenced papers published prior to 1900 (see Marx et al., 2014; Marx and Bornmann, 2014).

4. For detecting important historical publications newer than 1900, we selected the reference publication year time period 1901-1970 and removed all references with reference counts less than 100 for further analysis. The number of references decreased to n=226. A minimum reference count of 100 for the references, which appeared between 1900 and 1970, again results from the need to reduce the flood of references.

**2.2 Climate change research literature citing the marker paper**

In the RPYS-CO, we focused on the discovery of the greenhouse effect and the specific role of carbon dioxide. The new version of the CRExplorer allows the restriction of the cited references to only those which are co-cited with at least one selected cited reference using the menu items "Data" – "Retain Publications citing Selected Cited References". In a first step, we selected the marker reference in the table visualized by the CRExplorer. Note that also misspelled reference variants have to be considered, because in the case of earlier references their portion can not be neglected (Marx, 2011). Misspelled variants can be considered in two ways: (1) All misspelled variants are selected and the abovementioned menu item is chosen. (2) All misspelled variants are clustered and merged. Afterwards, the resulting unified cited reference is selected and the aforementioned menu item is chosen. Both possibilities yield the same RPYS-CO result. Since we used a dataset in this study with merged variants of cited references, we had to select only one reference.

Using the menue item of the CRExplorer in a second step, we reduced the references (n=10,932,050) of our publication set (n=222,060) to all references co-cited with the 1896 paper by Svante Arrhenius (n=1658). In contrast to the reference analysis of the literature in total, the co-cited references were not restricted by a minimum reference count (see above).

Svante Arrhenius was the first scientist who calculated how changes in the levels of carbon dioxide in the atmosphere could alter the surface temperature through the greenhouse effect. He predicted that emissions of carbon dioxide from the burning of fossil fuels were large enough to cause global warming. Interestingly, his prediction of a global temperature rise through doubling of the carbon dioxide concentration (about 5 degrees Celsius) is close to the predictions of recent climate models. Obviously, the many simplifications and mistakes of his calculation averaged out. Based on the growth rate observed at his time, Arrhenius did not expect a doubling of the atmospheric carbon dioxide concentration within the next 3000 years. The concentration increase currently observed (see: http://www.esrl.noaa.gov/gmd/ccgg/trends/) means a doubling within only two centuries. His paper can be seen as a cornerstone in the evolution of climate change research. In the first half of the 20$^{th}$ century, however, his theory was widely refused and accordingly, the paper was barely cited (until 1990, the number of citations per year was below ten).

By analyzing the co-citations of Arrhenius' paper as a marker reference, we investigate the discovery of the greenhouse effect and the specific role of carbon dioxide. This research topic marks the historical roots and origins of the current climate change research. By zooming from climate change research in total to the discovery of the greenhouse effect, we focus on literature specifically discussing the history of climate change research. We reveal the dicisive works in this context, including the forerunners and the papers taking up the impetus given by Arrhenius.

**3. Results**
**3.1 Historical roots of climate change research**

FIGs 1-2 show the results from the RPYS based on the climate change research literature in total. The spectrograms show the distribution of the number of cited references across their publication years within the two time periods 1686-1900 and 1901-1970. The earliest cited reference year visible in FIG 1 is 1686.The earliest reference publication year of the original

reference set is the year 1002, but between 1002 and 1686 there are no references with reference counts above the minimum of 10 citations.

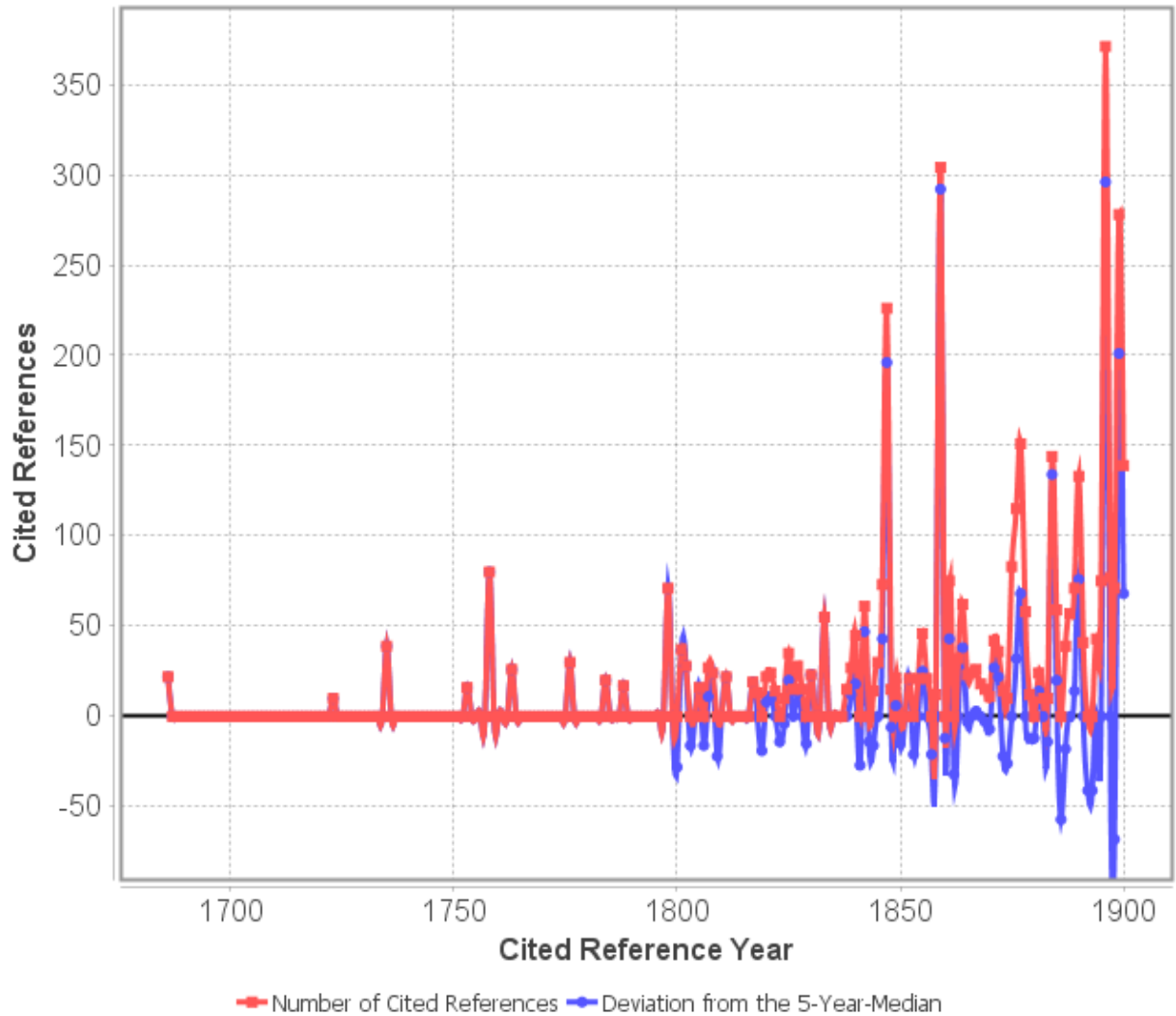

FIG 1: Annual distribution of cited references throughout the time period 1686-1900, which have been cited in climate change papers (published between 1980 and 2014). As a consequence of line smoothing the red curve crosses the zero line.

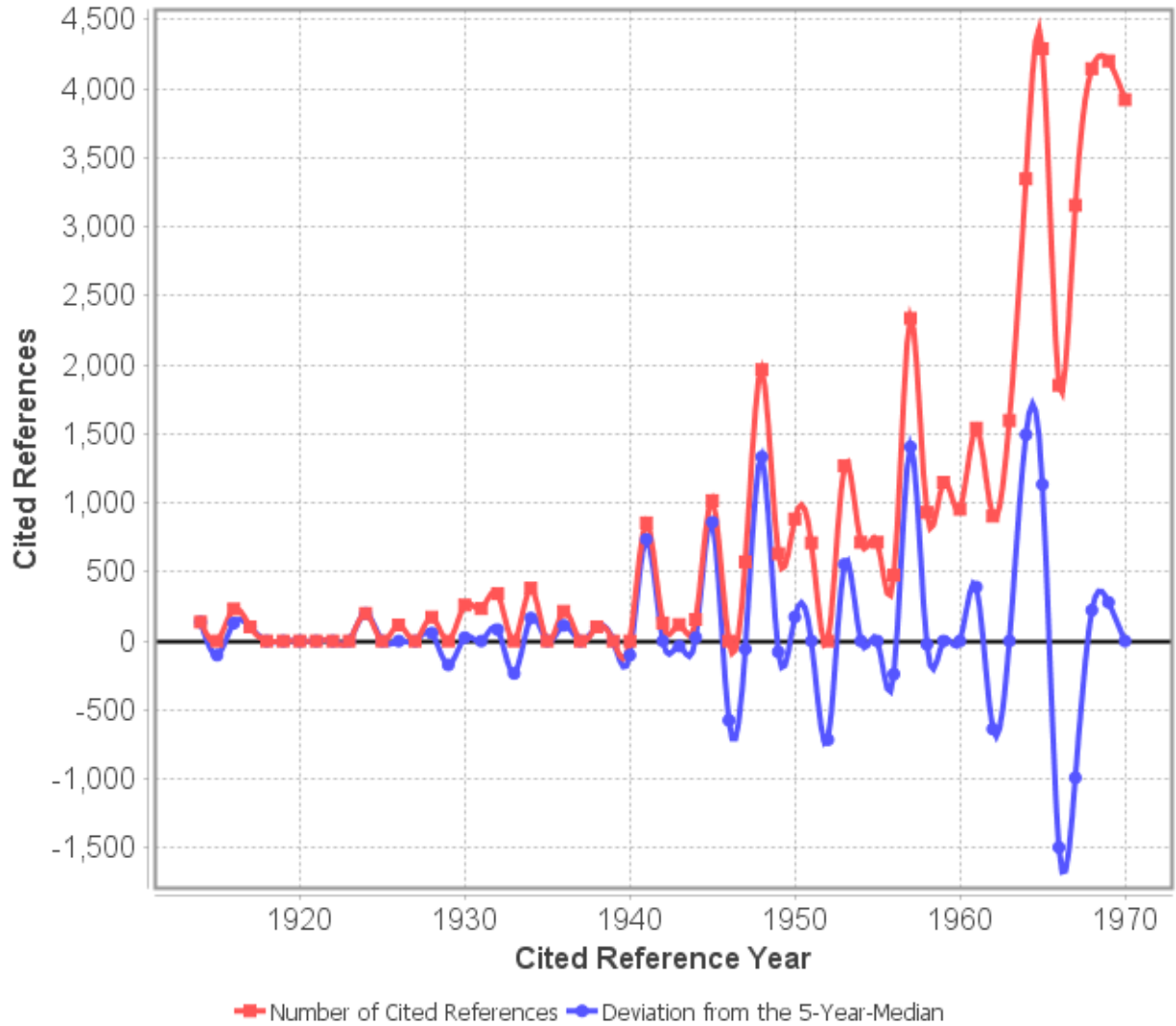

FIG 2: Annual distribution of cited references throughout the time period 1901-1970, which have been cited in climate change papers (published between 1980 and 2014). As a consequence of line smoothing the red curve crosses the zero line.

The red lines in FIGs 1-2 visualize the number of cited references per reference publication year. In order to identify those publication years with significantly more cited references than adjacent years, the deviation of the number of cited references in each year from the five-year median (median of the number of cited references in the two previous, the current, and the two following years: $t-2$; $t-1$; $t$; $t+1$; $t+2$) is also visualized (blue lines). The deviation from the five-year median (blue lines) provides a curve smoother than the curve showing absolute numbers (red lines). The blue line makes the distinct peaks more clearly visible. The more recent the references, the less distinct are the peaks, because increasingly more different references per year smooth both the number of cited references and the deviation from the median.

According to FIGs 1-2, there are pronounced peaks in the following 25 reference publication years: 1686, 1735, 1758, 1798, 1847, 1859, 1884, 1896, 1899, 1916, 1932, 1941, 1945, 1948, 1950, 1953, 1957, 1961, 1963, 1964, 1965, 1967, 1968, 1969, and 1970. TAB 1 lists the 35 cited references, which comprise the vast majority (or even the complete number) of the cited references of the corresponding reference publication years in FIGs 1-2. As the results in TAB 1 show, it is possible that there are two or more important referenced papers within one and the same peak (e.g. see the two cited references with RPY=1941 or the three cited references with RPY=1965). Short comments are added, which summarize the content of the cited works and (as far as possible) their relation to climate change research.

TAB 1: The most frequently cited references from specific reference publication years in FIGs 1-2, which have been cited by papers dealing with climate change. For each reference (CR), a sequential number (No), the corresponding reference publication year (RPY), and the number of cited references (NCR) within the climate change related publication set are listed. Furthermore, a short comment is added each, which explains the content of the cited work and its relation to climate change research.

| No  | RPY  | Reference / Comment | NCR |
|-----|------|---------------------|-----|
| CR1 | 1686 | E. Halley: An historical account of the trade winds, and monsoons, observable in the seas between and near the tropicks, with an attempt to assign the physical cause of the said winds. Philosophical Transactions of the Royal Society of London 16(179-191), 153-168 (1686). | 22 |
|     |      | *Edmond Halley (1656–1742) was an English astronomer, geophysicist, and meteorologist, well known for computing the orbit of the eponymous Halley's Comet. In the paper on trade winds and monsoons he published results from his Helenian expedition. He identified solar heating as the cause of atmospheric motions.* | |
| CR2 | 1735 | G. Hadley: Concerning the cause of the general trade-winds. Philosophical Transactions of the Royal Society of London 39(436-444), 58-62 (1735). DOI: 10.1098/rstl.1735.0014 | 24 |
|     |      | *The publication by George Hadley (1685–1768) describes an atmospheric circulation system that bears the authors name (Hadley Cell). This circulation system is intimately related to the trade winds, the tropical rainbelts, the subtropical deserts, and the jet streams.* | |
| CR3 | 1758 | C. Linnaeus: Systema naturæ per regna tria naturæ, secundum classes, ordines, genera, species, cum characteribus, differentiis, synonymis, locis 1 (10th ed.). Stockholm, Sweden: Laurentius Salvius, pp. [1–4], 824 pages (1758). | 80 |
|     |      | *The publication is one of the major publications of the Swedish botanist, zoologist, and physician Carolus Linnaeus (1707–1778) and introduced the Linnaean taxonomy. The first edition of the book was printed 1735. The tenth edition (1758) is considered the starting point of zoological nomenclature.* | |

12| CR4 | 1798 | T.R. Malthus: An essay on the principle of population. In Oxford World's Classics reprint. Printed for J. Johnson, in St. Paul's Church-Yard, London, UK (1798). | 71 |
|---|---|---|---|
| | | *The book was first published in 1798 under the alias Joseph Johnson. The author Thomas Robert Malthus (1766–1834) stated that sooner or later population will be checked by famine and disease, leading to what is known as a Malthusian catastrophe.* | |
| CR5 | 1847 | C. Bergmann: Über die Verhältnisse der Wärmeökonomie der Thiere zu ihrer Grösse. In: Göttinger Studien, 1. Abt., 595–708 (1847). | 206 |
| | | *The paper by the German biologist Carl Bergmann (1814–1865) describes what has been named Bergmann's rule: A principle that states that populations and species of larger size are found in colder environments, and species of smaller size are found in warmer regions.* | |
| CR6 | 1859 | C. Darwin: On the origin of species by means of natural selection, or the preservation of favoured races in the struggle for life. Murray, London, UK (1859). | 292 |
| | | *This seminal publication by Charles Darwin (1809–1882) is considered to be the foundation of evolutionary biology. The book attracted widespread interest upon its publication and generated scientific, philosophical, and religious discussion until present.* | |
| CR7 | 1884 | H.F. Blanford: On the connexion of the Himalaya snowfall with dry winds and seasons of drought in India. Proceedings of the Royal Society of London 37(232-234), 3-22 (1884). DOI: 10.1098/rspl.1884.0003 | 107 |
| | | *Henry Francis Blanford (1834–1893) was a British meteorologist who made long-term weather forecasts using the link between snow in the Himalayas and rainfall in the rest of India. Using this method, he was able to predict a deficient monsoon in 1885.* | |
| CR8 | 1896 | S. Arrhenius: On the influence of carbonic acid in the air upon the temperature of the ground. Philosophical Magazine and Journal of Science Series 5(41), 237-276 (1896). | 311 |
| | | *Svante Arrhenius (1859–1927) was the first scientist who calculated how changes in the levels of carbon dioxide in the atmosphere could alter the surface temperature through the greenhouse effect. He predicted that emissions of* | |
| | | *carbon dioxide from the burning of fossil fuels were large enough to cause global warming. His prediction of a global temperature rise through doubling the carbon dioxide concentration is close to recent predictions. Until around 1950, experts refused the hypothesis because they found that he had grossly oversimplified the climate system.* | |
| CR9 | 1899 | T.C. Chamberlin: An attempt to frame a working hypothesis on the cause of glacial periods on an atmospheric basis. Journal of Geology 7, 545-584, 667-685, 751-787 (1899). | 111 |
| | | *Thomas Chrowder Chamberlin (1843–1928) was an influential American geologist. In his 1899 paper he proposed the possibility that changes in climate could result from changes in the concentration of atmospheric carbon dioxide – thereby supporting the theory of Arrhenius (1896).* | |
| CR10 | 1916 | F.E. Clements: Plant succession: An analysis of the development of vegetation. Publication 242 (512 pages), Carnegie Institution of Washington, Washington, USA (1916). | 125 |
| | | *Frederic Edward Clements (1874–1945) was an American plant ecologist and pioneer in the study of vegetation succession.* | |
| CR11 | 1932 | G.T. Walker and E.W. Bliss: World weather V, Memoirs of the Royal Meteorological Society 4(36), 53-84 (1932). http://www.rmets.org/publications/classic-papers | 343 |
| | | *Sir Gilbert Thomas Walker (1868–1958) was a British physicist and statistician. He is best known for his groundbreaking description of the Southern Oscillation, a major phenomenon of global climate, for discovering Walker Circulation, and for greatly advancing the study of climate in general.* | |
| CR12 | 1941 | H. Jenny: Factors of soil formation: A system of quantitative pedology. Dover Publications Inc., New York, USA (1965). Originally published: McGraw-Hill, New York, USA (1941). | 354 |
| | | *The monograph by Hans Jenny is an advanced treatise on theoretical soil science and includes a study of soil-forming factors and processes of soil genesis. It is an extension of the first part of the course "Development and morphology of soils" at the College of Agriculture of the University of California.* | |
| CR13 | 1941 | M. Milankowic/Milankovitch: Canon of insolation and the ice-age problem (Kanon der Erdbestrahlung und seine Anwendung auf das Eiszeitenproblem). Royal Serbian Academy special publications, Section of Mathematical and | 352 |
| | | Natural Sciences 132, V33, 633 pages, Belgrade, Yugoslavia (1941). English translation by N. Pantic: Canon of Insolation and the Ice Age Problem. Alven Global, 636 pages (1998). | |
| | | *Milutin Milankowic/Milankovitch (1879–1958) explained how the earth's long-term climate changes (in particular the ice ages occurring in the geological past of the earth) are caused by changes in the position of the earth in comparison to the Sun, now known as Milankovitch Cycles.* | |
| CR14 | 1945 | H.B. Mann: Nonparametric tests against trend. Econometrica 13(3), 245-259 (1945). | 908 |
| | | *The paper by Henry B. Mann discusses tests of randomness against trends.* | |
| CR15 | 1948 | C.W. Thornthwaite: An approach toward a rational classification of climate. Geographical Review 38(1), 55-94 (1948). | 1078 |
| | | *The American climatologist and geographer Charles Warren Thornthwaite (1899–1963) devised a moisture based climate classification system (that is still in use worldwide) by monitoring the soil water budget.* | |
| CR16 | 1948 | H.L. Penman: Natural evaporation from open water, bare soil and grass. Proceedings of the Royal Society of London A - Mathematical and Physical Sciences 193(1032), 120-145 (1948). | 665 |
| | | *In the paper two theoretical approaches to the evaporation from saturated surfaces are outlined.* | |
| CR17 | 1950 | J.M. McCrea: On the isotopic chemistry of carbonates and a paleotemperature scale. Journal of Chemical Physics 18(6), 849-857 (1950). DOI: 10.1063/1.1747785 | 318 |
| | | *The temperature variation of the fractionation of oxygen in exchange reactions between dissolved carbonate and water and between calcite and water are calculated on theoretical grounds and checked experimentally.* | |
| CR18 | 1953 | S. Epstein et al.: Revised carbonate-water isotopic temperature scale. Geological Society of America Bulletin 64(11), 1315-1325 (1953). DOI: 10.1130/0016-7606(1953)64[1315:RCITS]2.0.CO;2 | 621 |
| | | *Samuel Epstein (1919–2001) explored the field of stable isotope geochemistry and studied natural variations in the isotopic abundances of hydrogen, carbon, oxygen, and silicon, with applications to archeology and climatology.* | |

| CR19 | 1957 | G.H. Hutchinson: Population studies: Animal ecology and demography. Cold Spring Harbor Symposia on Quantitative Biology 22, 415-427 (1957). Concluding remarks reprinted in: Bulletin of Mathematical Biology 53(1/2), 193-213 (1991). | 656 |
|---|---|---|---|
| | | *George Evelyn Hutchinson (1903–1991) is an American ecologist sometimes described as the "father of modern ecology". He contributed a mathematical theory of population growth to climate research.* | |
| CR20 | 1961 | H. Craig: Isotopic variations in meteoric waters. Science 133(346), 1702-1703 (1961). DOI: 10.1126/science.133.3465.1702 | 489 |
| | | *The paper reports how the relationship between deuterium and oxygen-18 concentrations in natural meteoric waters from many parts of the world has been determined with a mass spectrometer.* | |
| CR21 | 1961 | H. Stommel: Thermohaline convection with 2 stable regimes of flow. Tellus 13(2), 224-230 (1961). | 406 |
| | | *The paper deals with the thermohaline ocean circulation. In contrast to wind-driven currents, the thermohaline circulation is part of the ocean circulation, which is driven by density differences.* | |
| CR22 | 1963 | E.N. Lorenz: Deterministic non-periodic flow. Journal of the Atmospheric Sciences 20(2), 130-141 (1963). DOI: 10.1175/1520-0469(1963)020<0130:DNF>2.0.CO;2 | 493 |
| | | *Edward Norton Lorenz (1917– 2008) was an American mathematician, meteorologist, and a pioneer of chaos theory. He introduced the strange attractor notion and coined the term "butterfly effect", which is most important for the basic limits of weather forecasting.* | |
| CR23 | 1964 | W. Dansgaard: Stable isotopes in precipitation. Tellus 16(4), 436-468 (1964). | 1337 |
| | | *The paper is most-important for the reconstruction of the past climate based on ice core samples. Willi Dansgaard (1922–2011) was the first scientist to demonstrate that measurements of the trace isotopes deuterium and oxygen-18 in accumulated glacier ice could be used as an indicator of past climate.* | |
| CR24 | 1965 | W.C. Palmer: Meteorological drought. Research paper no. 45, U.S. Department of Commerce & Office of Climatology, Weather Bureau, February 1965 (58 pages). http://www.ncdc.noaa.gov/temp-and-precip/drought/docs/palmer.pdf | 767 |
| | | *The American meteorologist Wayne Palmer published the so-called Palmer drought index as a measurement of dryness based on recent precipitation and temperature. A highly topical work in view of the increasing dryness, e.g. in the western part of the USA.* | |
| CR25 | 1965 | J.L. Monteith: Evaporation and environment. Symposium of the Society for Experimental Biology 19, 205-234 (1965). | 637 |
| | | *The paper deals with evaporation of plants / leaves.* | |
| CR26 | 1965 | H Craig and L.I. Gordon: Deuterium and oxygen 18 variations in the ocean and the marine atmosphere. In: Stable isotopes in oceanographic studies and paleotemperatures. Editor: E. Tongiorgi, Spoleto, Italy (1965). | 556 |
| | | *This publication is important for measurements of the isotopes deuterium and oxygen-18 as an indicator of paleotemperatures and thereby for the reconstruction of the past climate.* | |
| CR27 | 1967 | S. Manabe and R.T. Wetherald: Thermal equilibrium of atmosphere with a given distribution of relative humidity. Journal of Atmospheric Sciences 24(3), 241-259 (1967). | 454 |
| | | *The "Manabe-Wetherald one-dimensional radiative-convective model" is seen as the first realistic atmospheric model, which considers the convection and radiation budget of the atmosphere.* | |
| CR28 | 1967 | R.H. MacArthur and E.O. Wilson: The theory of island biogeography. Princeton University Press (203 pages), Princeton, USA (1967). | 440 |
| | | *The authors developed a general theory to explain the facts of island biogeography. Their work provided a new framework to explain patterns in species diversity.* | |
| CR29 | 1968 | M.A. Stokes and T.L. Smiley: An introduction to tree-ring dating. University of Chicago Press (73 pages), Chicago, USA (1968). | 705 |
| | | *The monograph introduces the method of dendrochronology (tree-ring dating) based on the analysis of patterns of tree rings. One main area of application of the method is paleoecology (reconstruction of the past climate).* | |
| CR30 | 1968 | P.K. Sen: Estimates of regression coefficient based on Kendalls tau. Journal of the American Statistical Association 63(324), 1379-1389 (1968). DOI: 10.1080/01621459.1968.10480934 | 624 |
| | | *A simple and robust estimator of β based on Kendall's rank correlation tau is studied and presented.* | |

| CR31 | 1968 | G. Hardin: The tragedy of the commons. Science 162(3859), 1243-1248 (1968). DOI: 10.1126/science.162.3859.1243 | 410 |
|---|---|---|---|
| | | *The paper by Garrett Hardin is based upon an essay by a Victorian economist on the effects of unregulated grazing on common land and denotes a situation where individuals acting independently and rationally according to each's self-interested behaviour contrary to the common interests of the whole group.* | |
| CR32 | 1969 | J. Bjerknes: Atmospheric teleconnections from equatorial pacific. Monthly Weather Review 97(3), 163-172 (1969). DOI: 10.1175/1520-0493(1969)097<0163:ATFTEP>2.3.CO;2 | 634 |
| | | *Jacob Bjerknes helped toward an understanding of El Niño Southern Oscillation (ENSO). He suggested that a weakening of the east-west temperature difference can disrupt trade winds, resulting in increasingly warm water toward the east (see also Walker & Bliss, 1932).* | |
| CR33 | 1969 | S. Manabe: Climate and ocean circulation I: Atmospheric circulation and hydrology of earth's surface. Monthly Weather Review 97(11), 739-774 (1969). | 498 |
| | | *The paper outlines how the effect of the hydrology of the earth's surface can be incorporated into a numerical model of the general circulation of the atmosphere.* | |
| CR34 | 1969 | M.I. Budyko: Effect of solar radiation variations on climate of earth. Tellus 21(5), 611-619 (1969). | 458 |
| | | *From the analysis of observation data, the paper shows that variations of the mean temperature of the earth can be explained by the variation of solar radiation, arriving at the earth's surface.* | |
| CR35 | 1970 | J.E. Nash and J.V. Sutcliffe: River flow forecasting through conceptual models part I — A discussion of principles. Journal of Hydrology 10(3), 282-290 (1970). | 1332 |
| | | *The principles governing the application of the conceptual model technique to river flow forecasting are discussed. The necessity for a systematic approach to the development and testing of the model is explained and some preliminary ideas are suggested.* | |

The numbers of cited references presented in TAB 1 are substantially lower than the overall numbers of citations (times cited information of the WoS database records or number of citations based on the WoS cited reference search mode) of the corresponding papers. The main reason for this discrepancy is that a large portion of the references are cited by papers outside the climate change research literature. For example, the key paper by Arrhenius (1896) is also cited by works in geoscience dealing with the carbon cycle. Also, the database records of some relevant papers do not include an abstract and others do not include the search terms of our query in their title or abstract text. Therefore, such papers are not included in our publication set.

Note that the reference counts of all references within a specific reference publication year (i.e. references of the same age) are directly comparable among each other since all citing papers belong to the same research field: climate change research. Thus, the cited references originate in the same citation culture and it is not necessary to field-normalize the cited reference counts as it is conventional in bibliometrics (Waltman, 2016). The field-normalization is ensured by the first step of the RPYS: the selection of the publication set on which citation impact is measured target-oriented. However, the corresponding papers of the references presented in TAB 1 have been published throughout a large time period (1686-1970) within quite different publication and citation cultures. Thus, the reference counts in the table from different time periods are not comparable with each other. The average (and maximum) number of cited references per paper increases from past to present (Bornmann and Haunschild, 2016).

According to TAB 1, some of the most frequently cited early publications are most prominent publications of science in general. They are only loosely connected to climate change, in particular the publications by Carolus Linnaeus (CR3), Robert Malthus (CR4), and Charles Darwin (CR6). Their basic nature causes comparatively high citation rates also in the climate change research field. Linnaeus is seen as the father of modern taxonomy and is also considered as one of the founders of modern ecology, where climate plays a major role as driving force. Darwin as the creator of the evolutionary theory was concerned with climate as an important factor of natural selection. Malthus uncovered famine and disease (both often a result of climatic changes) as limiting factors for the growth of population.

Other early works are basic publications in specific disciplines like meteorology (Halley, CR1; Hadley, CR2) or biology (Bergmann, CR5) with a connection to climatology. Edmond Halley identified solar heating as the cause of atmospheric motions and George Hadley discovered that the atmospheric circulation system is intimately related to monsoons and trade winds, the tropical rainbelts, the subtropical deserts, and the jet streams. Carl Bergmann introduced what has been named Bergmann's rule: A principle that states that populations and species of larger size are found in colder environments, and species of smaller size are found in warmer regions.

These early publications do not mark the beginning of the actual climate change research. The scientific discovery of climate change did not begin before the early 19th century when the ice ages and the greenhouse effect were discovered. The first publication, which can be seen as the origin of modern research on climate change, was published 1896 by Arrhenius (CR8). Thus, this paper has been used as marker reference for the co-citation analysis in section 3.2 below.

The most frequently cited papers published within the first half of the 20th century are seminal papers in meteorology (Walker and Bliss, CR11), agronomy (Jenny, CR12; Penman, CR16), and climatology (Thornthwaite, CR15). Gilbert Thomas Walker is best known for his groundbreaking description of the *Southern Oscillation* and for discovering the *Walker Circulation*. The most frequently cited papers, which appeared in the beginning of the second half of the 20th century, are related to various fields of geoscience: The publications by Stommel (CR21), Bjerknes (CR32), and Manabe (CR33) deal with oceanic currents, which are major phenomena of global climate. The paper by Manabe and Wetherald (CR27) introduced the first realistic atmospheric model, which considers the convection and radiation budget of the atmosphere. The papers by McCrea (CR17), Epstein et al. (CR18), and Craig and Gordon (CR26) are most important for isotope based dating of ice cores and of all kinds of sediments, whereas the paper by Stokes and Smiley (CR29) gives an introduction to tree-ring dating (dendrochronology). These papers document the importance of research on paleoclimate, both for understanding the earth's climatic system and for future climate predictions.

The publication by Budyko (CR34) refers to the variations of solar activity as a natural climate factor, which is discussed until present. Blanford (CR7), Penman (CR16), Palmer (CR24), and Monteith (CR25) deal with evaporation and drought, which are decisive factors for agriculture. These are highly topical works in view of the increasing dryness, e.g. in the western part of the USA. The publications by Hutchinson (CR19) and MacArthur and Wilson (CR28) are biosphere related and deal with animal ecology and plant formation or succession. The fundamental work by Milankowitch (CR13) on the causes of glacial periods and by Lorenz (CR22) on chaos theory (see the butterfly effect) are cornerstones of science and have a strong connection to climate change and meteorology, respectively.

Two highly cited pre-1971 publications are the papers by Dansgaard (CR23) and Nash and Sutcliffe (CR35). The paper by Dansgaard (CR23) is most important for the reconstruction of the past climate based on ice core samples. Willi Dansgaard was the first scientist who demonstrated that measurements of the trace isotopes deuterium and oxygen-18 in accumulated glacier ice could be used as an indicator of past climate. The paper by Nash and Sutcliffe (CR35) is one of the first publications about climate change vulnerability. The paper by Hardin (CR31) is an exception in TAB 1: "The tragedy of the commons" concept is often cited in connection with sustainable development, meshing economic growth and environmental protection, as well as in the debate of global warming.

### 3.2 Climate research literature citing the marker paper

In the RPYS-CO we substantially reduced the climate change research publication set by using co-citation analysis. FIG 3 shows the spectrogram from the RPYS-CO using the references co-cited with Arrhenius (1896). By definition, the peak associated with this reference is dominating the results. There are two distinct peaks corresponding to references prior to 1896 resulting from the works of forerunners, and six peaks with reference publication years later than 1896. The CRExplorer reveals that the broader peak after 1896 includes three frequently cited papers published between 1897 and 1899. This peak is followed by pronounced peaks corresponding to the reference publication years 1908, 1924, 1938, and 1956/1957 (when modern climate change research began). In order to demonstrate the potential of a more detailed analysis with the CRExplorer, a spectrogram is included in FIG 3 which is only based on the cited reference years 1800-1850. We zoomed into this time period for an in-depth analysis of the citation pattern of the works of the French mathematician and physicist J.B.J. Fourier, which are sometimes discussed and cited unsatisfactorily in historical overviews on the history of climate change research (see Fleming, 1998, 1999).

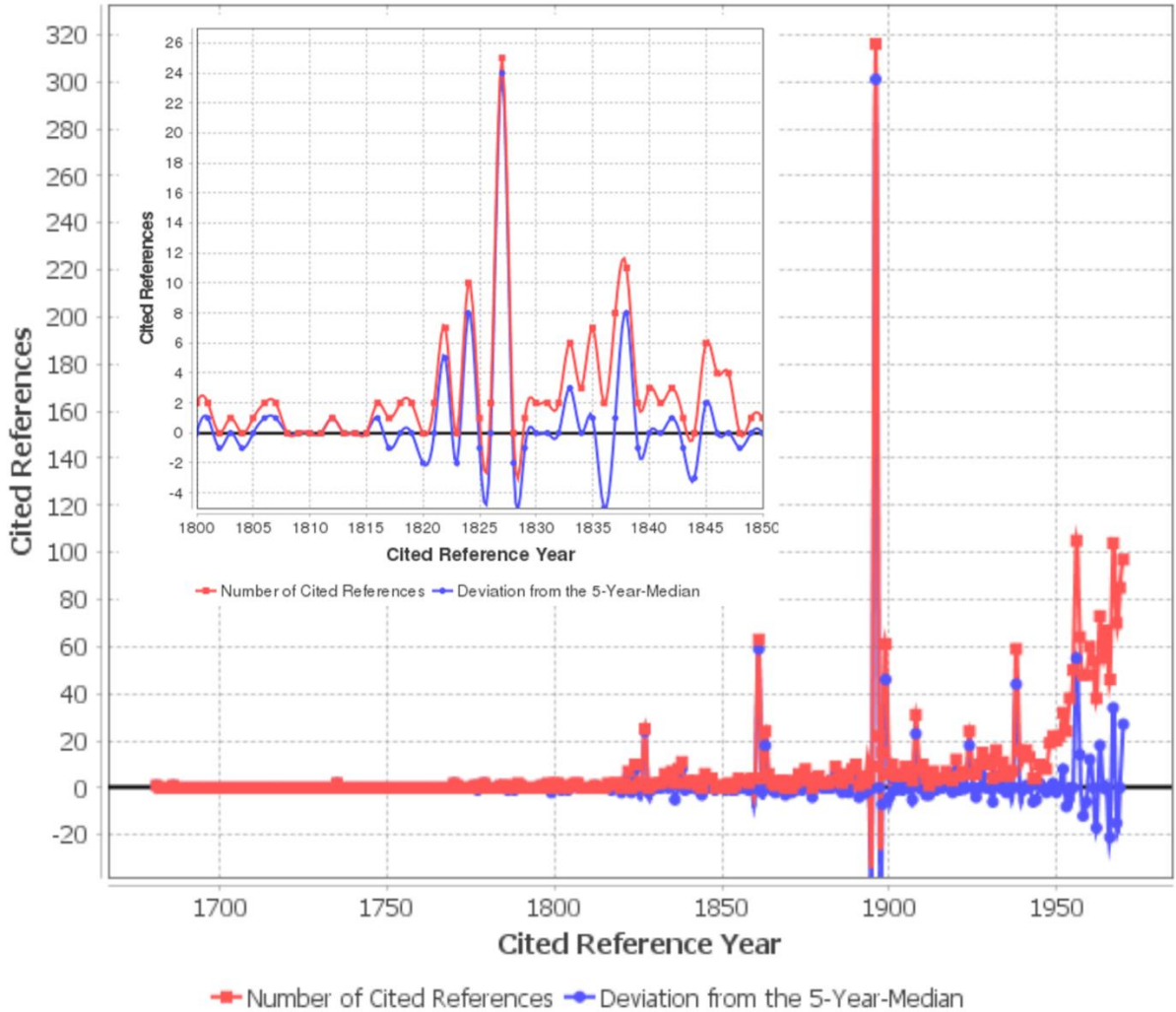

FIG 3: Annual distribution of cited references which are co-cited with Arrhenius (1896) throughout the time period 1686-1970. The curves in the top left corner focus on the time period 1800-1850 as one part of the overall figure. As a consequence of line smoothing the red curves crosses the zero line.

The cited references underlying the peak reference publication years across the complete time period 1686-1970 in FIG 3 are listed in TAB 2. Due to the narrower focus on the discovery of the greenhouse effect and the specific role of carbon dioxide, there are only two references, which appear in TAB 1 as well as in TAB 2: the 1896 paper by Arrhenius and the 1899 paper by Chamberlin.

TAB 2: The most frequently cited references from specific reference publication years in FIG 3, which have been co-cited with Arrhenius (1896). For each co-cited reference (COR), a sequential number (No), the corresponding reference publication year (RPY), and the number of co-cited references (NCOR) within the climate change related publication set are listed. Furthermore, a short comment is added each, which explains the content of the cited work and its relation to climate change research. The marker reference (Arrhenius, 1896) is shaded in gray.

| No | RPY | Reference / Comment | NCOR |
|---|---|---|---|
| COR1 | 1824* | J.B.J. Fourier: Remarques Générales sur les Températures du Globe Terrestre et des Espaces Planétaires. Annales de Chimie et de Physique 27, 136–167 (1824). | 25 |
| COR2 | 1827 | J.B.J. Fourier: Mémoire sur les températures du globe terrestre et des espaces planétaires. Mémoires de l'Académie Royale des Sciences 7, 569–604 (1827). | 20 |
|  |  | *The French mathematician and physicist Jean Baptiste Joseph Fourier (1768-1830) determined that the earth is warmer than expected according to his estimation. He attributed this to the phenomenon, that the earth's atmosphere is transparent for solar radiation, but not for the infrared radiation produced by the warming of the atmosphere and the ground. Thereby, Fourier discovered the (natural) greenhouse effect. His 1827 paper is a reproduction of is 1824 work, which in turn is based on his magnum opus of 1822 on the analytical theory of heat.* |  |
| COR3 | 1861 | J. Tyndall: On the absorption and radiation of heat by gasses and vapours, and on the physical connection of radiation, absorption, and conduction. Philosophical Magazine Series 4, 22(146-147), 169-194, 273-285 (1861). | 62 |
| COR4 | 1863 | J. Tyndall: On the radiation through the earth's atmosphere. Philosophical Magazine Series 4, 25(167), 200-206 (1863). | 12 |
|  |  | *The Irish physicist John Tyndall (1820-1893) was the first to measure the radiant heat (infrared) absorptive powers of greenhouse gases, including carbon dioxide and water vapor. He suggested that changes in the* |  |
| | | *concentration of these gases could result in climatic change. His contributions have been published in a variety of papers (around 1855-1875) with the 1861 paper as the most frequently cited work.* | |
| | | | |
| COR5 | 1896 | S. Arrhenius: On the influence of carbonic acid in the air upon the temperature of the ground. Philosophical Magazine and Journal of Science Series 5(41), 237-276 (1896). | 311 |
| | | *The marker reference (see TAB 1).* | |
| COR6 | 1897 | T.C. Chamberlin: A group of hypotheses bearing on climatic changes. Journal of Geology 5, 653-683 (1897). | 17 |
| COR7 | 1898 | T.C. Chamberlin: The influence of great epochs of limestone formation upon the constitution of the atmosphere. Journal of Geology 6, 609-621 (1898). | 13 |
| COR8 | 1899 | T.C. Chamberlin: An attempt to frame a working hypothesis on the cause of glacial periods on an atmospheric basis. Journal of Geology 7, 545-584, 667-685, 751-787 (1899). | 55 |
| | | *Thomas Chrowder Chamberlin (1843–1928) was an influential American geologist. In his 1899 paper he proposed the possibility that changes in climate could result from changes in the concentration of atmospheric carbon dioxide – thereby supporting the theory of Arrhenius (1896).* | |
| COR9 | 1908 | S. Arrhenius: Worlds in the making: the evolution of the universe. Harper & Brothers Publishers, New York & London, USA and UK (264 pages). | 35 |
| | | *This book was directed at a general audience and already appeared in 1906 in Swedish (Världarnas utveckling) and 1907 in German (Das Werden der Welten). Arrhenius concluded that the human emission of $CO_2$ would be strong enough to prevent the world from entering a new ice age, and also that a warmer earth would be needed to feed the rapidly increasing population.* | |
| COR10 | 1938 | G.S. Callendar: The artificial production of carbon dioxide and its influence on temperature. Quarterly Journal of the Royal Meteorological Society 64, 223-237 (1938). | 54 |
| COR11 | 1949 | G.S. Callendar: Can carbon dioxide influence climate? Weather 4, 310-314 (1949). | 11 |
| | | *The British engineer Guy Stewart Callendar (1898-1964) developed the first complete theory of climatic change and stated in 1938 that carbon dioxide caused the warming trend of the preceding decades. He presented evidence that both temperature and the $CO_2$ level in the atmosphere had been rising over the past half-century. Callendar attempted to revive Arrhenius's greenhouse-effect theory and wrote about the potential effects of anthropogenic $CO_2$ emissions for a period of nearly three decades (1938-1961).* | |
| COR12 | 1956 | G.N. Plass: The carbon dioxide theory of climatic change. Tellus 8(2), 140-154 (1956). | 20 |
| COR13 | 1956 | G.N. Plass: Effect of carbon dioxide variations on climate. American Journal of Physics 24, 376-387 (1956). | 10 |
| COR14 | 1956 | G.N. Plass: The influence of the 15-micron carbon dioxide band on the atmospheric infrared cooling rate. Quarterly Journal of the Royal Meteorological Society 82, 310-324 (1956). | 10 |
| | | *The Canadian physicist Gilbert Norman Plass (1920-2004) calculated the transmission of radiation through the earth's atmosphere and predicted that doubling the level would bring a 3-4°C rise. He was the first to use a computer for climate modelling. Plass expected that human activity would raise the average global temperature at the rate of 1.1 degrees C per century if emissions would continue at the 1950s rate.* | |
| COR15 | 1957 | R. Revelle and H.E. Suess: Carbon dioxide exchange between atmosphere and ocean and the question of an increase of atmospheric $CO_2$ during the past decades. Tellus 9, 18-27 (1957). | 26 |
| | | *Roger Revelle and Hans E. Suess remarked that mankind had embarked upon a "large-scale geophysical experiment" as a result of the tremendous utilization of fossil fuels by the industrialized society. However, the theory of an anthropogenic greenhouse effect was widely disputed or ignored until Charles David Keeling started in 1958 to measure the (steadily increasing) concentration of carbon dioxide in the earth's atmosphere at Mauna Loa (Hawaii) and until scientists increasingly predicted global warming in the 1970s.* | |

\* The (correct) reference publication year 1824 was mostly confused with 1924 (17 out of 25 references). A comparison of the Fourier (1924) reference within selected WoS database records with the corresponding references of the original papers revealed that this mistake is caused by the database producer and not by the citing authors.

The thematic focus by using the marker reference (COR5) in the RPYS-CO initially reveals the works of two forerunners: (1) The French mathematician and physicist Jean Baptiste Joseph Fourier (COR1-2) is widely seen as the first who detected the greenhouse effect. (2) The Irish physicist John Tyndall (COR3-4) was the first to measure the radiant heat (infrared) absorptive powers of greenhouse gases, including carbon dioxide. Further peaks in the spectrogram can be explained as follows: The American geologist Thomas C. Chamberlin (COR6-8) proposed the possibility that changes in climate could result from changes in the concentration of atmospheric carbon dioxide. The British engineer Guy Callendar (COR10-11) stated that carbon dioxide caused the warming trend of the preceding decades. The Canadian physicist Gilbert N. Plass (COR12-14) calculated the transmission of radiation through the earth's atmosphere and predicted that doubling the level would bring a 3-4°C rise. Roger Revelle and Hans E. Suess (COR15) remarked for the first time that mankind had started to embark upon a "large-scale geophysical experiment" as a result of the tremendous utilization of fossil fuels by the industrialized society.

## 4. Discussion

The first analysis of this study is based on a large and carefully searched publication set of 222,060 papers published between 1980 and 2014, which are dealing with research on climate change. The references cited therein were extracted and analyzed with regard to publications, which appeared prior to 1971 and are still cited most frequently. We identified the most pronounced peaks in the RPYS spectrogram and presented 35 publications together with comments on their content and relation to climate change. Thus, we focused on a very limited number of top-cited references showing the driving forces in the very heterogeneous field of climate change research. Even these 35 top references are too numerous to be discussed here very detailed.

The 35 publications include fundamental early works of the 19$^{th}$ century science and cornerstones of science that are closely related to climate change. The most frequently cited publications, which appeared in the beginning of the second half of the 20$^{th}$ century, are mostly published in various fields of geosciences including meteorology (atmospheric teleconnections and oceanic currents). Furthermore, works on agriculture, paleoclimatology (in particular isotope based dating of ice cores and sediments) and on climate change vulnerability play a major role in climate change research. Two highly cited pre-1971 publications are Dansgaard (CR23) and Nash (CR35). Dansgaard (CR23) is most-important for the reconstruction of the past climate based on ice core samples. Nash (CR35) is one of the first papers around climate change vulnerability.

The most frequently cited publications from both RPYS and RPYS-CO analyses (listed in TAB 1 and TAB 2) mirror the literature selection and accentuation of the overall climate change research community (i.e. the authors of the citing papers in our publication set). RPYS (including RPYS-CO) is a quantitative method; the results of the analysis can be compared with a qualitative approach by which experts in the field select and describe the important publications in a literature overview. As an example, we analyzed the "Historical overview of climate change" by Le Treut et al. (2007), a contribution to the Fourth Assessment Report of the Intergovernmental Panel on Climate Change (IPCC). This overview cites 13 of the 46 references listed in TAB 1 and 4 of the 15 references listed in TAB 2. For several authors of the referenced publications in TAB 1 (e.g. Manabe) we find that Le Treut et al. (2007) prefer to cite the more recent papers of these authors rather than their earlier papers published prior to 1971 and revealed by RPYS and RPYS-CO, respectively. The different focus explains most of the differences between the qualitative (historical literature overview) and the quantitative (bibliometric) approach.

Another reason for the differences between both approaches might be that importance does not always manifest in citations. A good example for important publications with comparatively low citation impact is the contribution by Charles David Keeling (e.g. Keeling, 1958, 1960), who started in 1958 to measure the concentration of carbon dioxide in the earth's atmosphere at Mauna Loa (Hawaii). His data collection is the longest continuous record of atmospheric carbon dioxide in the world, showing that carbon dioxide levels were rising steadily (Keeling Curve) – thereby revealing the possibility of global warming. The contributions of this pioneer have been published in a variety of papers since 1958 and appear as cited references in our dataset, but not as pronounced peaks in the spectrogroms of FIG 2 and FIG 3.

Despite Keeling's important contributions to the climate change research, his publications are not referenced extraordinarily high as explicit (formal, reference based) citations in papers on climate change. They seem to be cases of "obliteration by incorporation" (McCain, 2012), which affects seminal works that are rapidly absorbed into the body of scientifc knowledge. In the case of Keeling, the data collection for the curve bearing his name extended throughout decades; the idea behind and the result of the measurements is not presented in a single paper.

The second analysis of this study focusses on the discovery of the greenhouse effect and the specific role of carbon dioxide. For narrowing down the perspective from climate change research in total to the discovery of the greenhouse effect, we investigated literature specifically discussing the history of climate change research related to this effect and zoomed into the (much smaller) reference set cited therein. The data for the analysis is based on co-citations of the seminal work by Arrhenius (1896). We assumed that the relevant early works (if at all) are cited alongside with this prominent paper. It is hardly imaginable that the Arrhenius paper is not cited in the context of the history of climate change research in connection with carbon dioxide. We applied RPYS-CO on this more specific reference set and revealed a multitude of early works relevant for the history of climate change research. With the exception of Arrhenius (CR8) and Chamberlin (CR9), these papers do not appear as distinct peaks in the RPYS (the first analysis of this study).

We consulted the historical overviews of experts, in particular those of James Fleming (1998, 1999), Spencer Weart (1997, 2008), and Mike Hulme (2009). In his book entitled "Historical perspectives on climate change" Fleming states: "Most people writing on the history of the greenhouse effect merely cite in passing Fourier's descriptive memoir of 1827 as the 'first' to compare the heating of the earth's atmosphere to the action of glass in a greenhouse. There is usually no evidence that they have read Fourier's original papers or manuscripts (in French) or have searched beyond the secondary sources" (p. 55). According to Fleming (1999), "Svante Arrhenius (1896) began the practice of citing Fourier's 1827 reprint as the first mention of the greenhouse effect" (p. 74). Weart (2008) mentiones (but not cites) Fourier's papers and accentuates the contribution by Arrhenius, whereas Hulme (2009) emphasizes Tyndall's works.

In order to demonstrate the potential of the RPYS-CO approach for reconstructing the emergence and evolution of research dealing with the greenhouse effect, we focused on the time period 1800-1850 to reveal the references related to Fourier. Although barely cited, his 1822 and 1824 works appear as distinct peaks and can easily be identified. The preceding contributions, however, do not appear as pronounced peaks in the spectrogram but can be identified via the reference list established by the CRExplorer. The CRExplorer enables an in-depth analysis as requested by Fleming (1998): "Obviously, those seeking to understand the history of terrestrial temperature research must look well beyond the secondary literature and well before 1827" (p. 64). Also, a basic problem of historical overviews becomes evident when the reference analysis is applied (Fleming, 1998): "Simple claims about the 'discovery' of the greenhouse effect are impossible to sustain" (p. 64).

We cannot expect that RPYS and RPYS-CO deliver similar results in this study: The first analysis reveals the early works most relevant for a multitude of disciplines active in climate change research and presents only 35 prominent out of almost 11 Million cited references. The second analysis focuses on the discovery of the greenhouse effect and the specific role of carbon dioxide. Also, only the historically oriented climate change related publications are considered here. We suggest to further examine the latter approach in the form of RPYS-CO.

**5. Conclusions**

The results of this study demonstrated that RPYS reveals the publications which are most important for the evolution of the climate change research field. Hardly any science historian who has published an historical overview mentions (and cites) all relevant literature (to the best of our knowledge). According to a WoS based cited reference search, for example, there are only two papers concurrently citing Fourier and Tyndall and only three papers concurrently citing Fourier and Arrhenius (books presumably contain more co-citations). RPYS delivers a detailed list of potentially important publications which might be considered in an historical overview. The advantage of this method is that the seminal papers are detected on the basis of the references which are cited by the relevant community without any further assumptions.

Since importance is not always reflected by high citation counts, the reconstruction of the history of a specific research field is not possible without expert knowledge.The specific role of the cited papers in the historical context can only be determined by experts. Thus, the results of this study might serve scientists who are active in the field of the history of climate change as interesting material from a quantitative perspective. This study is an invitation to combine both approaches (expert knowledge and bibliometrics) for meaningful and informative historical analyses.

2929

**Acknowledgements**

The bibliometric data used in this paper are from an in-house database developed and maintained by the Max Planck Digital Library (MPDL, Munich) and derived from the Science Citation Index Expanded (SCI-E), Social Sciences Citation Index (SSCI), Arts and Humanities Citation Index (AHCI) provided by Thomson Reuters (Philadelphia, Pennsylvania, USA).